\documentclass{svcyclop}

\usepackage{graphicx}    
\usepackage{url}
\usepackage{color}

\usepackage[font=footnotesize]{subfig}

\begin{document}

\title{Analysis and Visualization of Dynamic Networks}

\author{
Faraz Zaidi\inst{1}
\and 
Chris Muelder\inst{2}
\and
Arnaud Sallaberry\inst{3}
} 

\institute{College of Computing and Information Sciences, Karachi Institute of Economics and Technology (KIET), Karachi, Pakistan\\							
            \url{faraz@pafkiet.edu.pk}
    \and
			Computer Science department, University of California at Davis, Davis, California, U.S.A\\
            \url{muelder@cs.ucdavis.edu}
    \and
    		Computer Science department, LIRMM - Universit\'e Montpellier 3, Montpellier, France\\
          	\url{arnaud.sallaberry@lirmm.fr}
      }    	
\maketitle

\section{Synonyms}
Temporal Networks or Graphs, Longitudinal Network Analysis, Evolving Networks or Graphs, Time-varying Graphs, Time-Stamped Graphs, Visual Analytics, Visual Data Mining, Information Visualization, Graph Mining, Network or Graph Visualization

\section{Glossary}

\begin{flushleft}\textbf{Network or a Graph:} A mathematical structure to represent objects and their interactions. Objects are represented by Nodes or Vertices (often denoted by a set $V$) and interactions are represented by Links or Edges (often denoted by a set $E$). Mathematically, a graph $G$ is defined as a tuple $G(V,E)$. Mathematicians use the term Graph whereas scientists from other disciplines usually use the term Network to refer to the same concept. Throughout this text, we use these terms interchangeably.
\end{flushleft}

\begin{flushleft}\textbf{Social Network:} A network where objects represent people and their interactions represent some sort of relationship among people. For example, two individuals may be connected to each other if they have studied at the same school, or play for the same football team.
\end{flushleft}

\begin{flushleft}\textbf{Clusters:} A group of nodes (representing objects) that are densely connected to each other and sparsely connected to other nodes in the network. Formally, a clustering of a static graph $G=(V,E)$ is defined by a set $C$ of subsets of $V$: $C=\{c_1,c_2,...,c_{l}\}$ such that $V=c_1 \cup c_2 \cup ... \cup c_l$. 
\end{flushleft}

\begin{flushleft}\textbf{Small World Network:} A graph with two characteristic properties. The average path length i.e.\ the number of nodes needed to traverse from one node to another on average is low, as compared to an equivalent size random graph. The second characteristic is the high transitivity among nodes i.e.\ many sets of three nodes are connected to each other with three vertices.
\end{flushleft}

\begin{flushleft}\textbf{Scale Free Network:} A graph whose degree distribution follows a power law where the power law coefficient is usually between [2,3]. In other words, this means that most nodes nodes have only a few connections (low degree) and few nodes have many connections (high degree) in the network.
\end{flushleft}


\section{Definition}

Network Science has emerged as an interdisciplinary field of study to model many physical and real world systems. A network, although consists of only a set of nodes and edges, but is a very powerful structure to represent a wide variety of systems such as people related through social relations, airports related through flights and computers connected through internet. The world we live in, increasingly becomes a collection of such networks and scientists from various disciplines are combining efforts to develop sound theories and fundamental concepts governing this new and exciting field of study. 

A sub-part of Network Science which has attracted lot of attention, practical applications and research interest is Social Network Analysis (SNA). SNA focuses on the relationships and the interconnected behaviour of different entities such as objects, people and organizations. A more specialized area associated with SNA is the study of dynamic behaviour of networks formally known as Dynamic Network Analysis (DNA). More often than not, social networks (and even networks in general), exhibit structural changes over time i.e.\ addition or deletion of nodes or edges. For example, social relations are a function of time as they appear and disappear with respect to social events taking place in the society. Similarly studying air transportation networks or data packet traffic over the internet, all are examples of dynamic networks where time plays an important role in the evolution, analysis and understanding of the entire network.

Research on Dynamic networks ranges from analytical and algorithmic models studying their evolution processes to the study of specialized topics such as the role of individuals in epidemics of infectious diseases. The study of these networks helps predict the behaviour of many natural and real world systems, how individuals or groups of individuals influence, control and mould the shape of the entire network.

Along with methods to analyze networks, visualization has become an integral component of this research area \cite{freeman04}. The visual representation of these networks exploit the human cognitive and perceptual capabilities to build hypothesis, validate theories, and create a better understanding of networks around us.

Formally, we can define a dynamic network as a network which undergoes structural changes over time. The analysis and visualization of these networks is the study of algorithms, methods, tools and techniques which help us understand these networks and extract applicable knowledge from them. The study of Dynamic Networks forms a new and cross disciplinary area of study with research opportunities and applications in many diverse fields. 

\section{Introduction}

Networks are all around us. Wherever we look, we can find an interconnected web of objects participating in an interdependent system. Traditional science studies a system of similar objects by modelling an individual and then generalizing its behaviour for the entire system. Network science is different, as its bases are in the holistic nature of systems that cannot be modeled as individuals but only through the study and analysis of interconnectedness. This holistic approach helps in understanding the complex behaviour of real world systems and allow us to predict and forecast how these networks evolve over time. The dynamic behaviour of many of these networks has made this field both interesting and challenging for researchers.  Subsequently this focus area of networks has been dubbed Dynamic Network Analysis (DNA).

Dynamic network analysis is the study of change occurring in networks with the passage of time \cite{moody05}. These dynamics occur due to processes either inherent to the system or through some external change processes forced into the system. An example of an inherent process for a a social network would be aging which results in death and thus change in social structure, simply due to the passage of time. In contrast, an external change process would be the development of a new airport in a city or the introduction of a technologically advanced hub in a communication network which induces structural change in these types of networks. The granularity of time with which change occurs depends on the system being modeled and the context in which it is studied. For example, the study of air traffic networks is meaningful in minutes and hours whereas the data packet traffic in communication networks is usually studied in micro and nano seconds.

People have only recently started to realize the potential of dynamic analysis and look closely at how it can be useful. Often, the temporal dimension has been completely ignored in earlier studies. The reasons for this fairly late thrust in this area (as compared to early work such as sampson's monastery study in 1969) is the immaturity of the field itself and the unavailability of reliable and substantial time-stamped datasets. Along with other datasets, the exponential growth of online social networks has boosted the availability of large, accurate and complete network datasets that include a temporal dimension, and have facilitated research in this area. This availability of dynamic network datasets has helped scientists move from theoretical framework to a more applied approach where they can actually formulate problems, test and validate their solutions pertaining to dynamic network analysis.

DNA is applied in scenarios where temporal ordering of events, temporal duration of processes or the rate of change in interactions taking place in a network are important. For example, researchers have studied the network of romantic relationships among school students to understand and model how sexually transmitted diseases spread in a network. A static view of the entire network fails to exhibit relationship changes over a time period. A partial view of the network of any specific day consists of mostly disconnected pairs of nodes as only a negligible number of students are involved in multiple relationships at the same time \cite{moody05}. For these kinds of networks, questions such as when and in what order did the relationships change, how frequently an individual changes a partner, all become significant thus enforcing the necessity of methods for the analysis of dynamic networks. Other examples of networks with dynamic behavior commonly found include data communication networks, social communication networks, transportation networks, ecological networks and biological networks. 

DNA differs from classical SNA as it studies the change phenomena rather than group level coherence, attribute level features and optimization issues. Questions related to the efficiency and stability of the network become more pertinent than the structural properties and organization of the network. As suggested by M. Trier, dynamic network analysis is `more concerned with the activity of actors and their relationships, as compared to static analysis which concentrates on structural issues of a network' \cite{trier08}. An important point to note is that dynamic network analysis does not replace static network analysis but it addresses a different set of issues altogether and has been found very useful in many application areas. 

Visualization of networks is an integral component of the field of SNA \cite{freeman04} and especially DNA as it helps to visually comprehend the dynamics taking place in a network. Real world systems can exhibit temporal dynamics in a number of ways. Visualizing these dynamics is an active area of research and several different methods have been proposed to solve a variety of real world problems. Consequently, methods that go beyond traditional SNA are needed to cater the growing requirements of DNA. Network analysis and visualization tools such as Pajek, UCINET, Gephi, Tulip and SoNIA (Social Network Image Animator) are all contributing towards the development of methods to support dynamic network analysis and visualization along with traditional SNA.


\section{Historical Background}

The history of static SNA dates back to 1933 from the field of sociology. Jacob Moreno \cite{moreno34}, a psychologist, used a Sociogram (which is now known as network) to represent what the interpersonal structure of a group of people looks like. He studied an epidemic of runaways at a school where he concluded that it was the structural positioning of students in the social network that caused students to runaway. Early work in network analysis with temporal data available is by T. M. Newcomb \cite{newcomb61} who for a period of sixteen weeks studied the acquaintance process and possible developing friendships in a group of male students who shared the same house. The first attempt to use visual analysis of dynamic networks was by Samuel F. Sampson as part of his Ph.D. thesis in 1968 \cite{sampson68}. Famously known as Sampson's monastery study, Sampson studied the evolution process of community structures in a New England monastery by taking several snapshots of the same network at different time intervals for visual analysis.

Recent interest in the field of SNA and subsequently DNA, was sparked by the milestone papers of Watts and Strogatz \cite{watts98}, and  Albert and Barabasi \cite{barabasi99} where they studied the properties of small world and scale free networks. These discoveries were made in very diverse fields such as collaboration network of film actors and neural network of the worm Caenorhabditis elegans, attracting researchers from other disciplines and triggering new research horizons across many disciplines.

The study of dynamic networks present a new and challenging area of research with its own set of problems. Domain experts relish methods that can help them visualize changes in a network resulting in better understanding and new discoveries from their network data. Although the field is in its infancy, but promises a lot for the future as more researchers focus on dynamic network analysis and visualization. Specialized journals (IEEE Network, IEEE Transactions on Visualization and Computer Graphics, Computer Graphics Forum, Information Visualization, Social Network Analysis and Mining, Journal of Social Structure) and conferences (such as IEEE VAST, IEEE InfoVis, EuroVis, IEEE/ACM ASONAM, IEEE SocialCom2011) are fast becoming standard platforms to share knowledge and encourage collaborative research in this area. 

\section{Dynamic Networks: Analysis and Visualization}

Research in the area of dynamic networks can be broadly categorized into two partially overlapping categories: The `Analytics' part and the `Visualization' part. These two categories are overlapping because many analytical methods are used in conjunction with visualization techniques to facilitate the overall process of interactive extraction of knowledge. Similarly many visualization techniques are used as a preprocessing step in the analysis phase to complement analytical methods. Before we discuss these two categories, we will give a mathematical definition of dynamic graphs followed by their general classification methods.

\subsection{Mathematical Formulation for Dynamic Networks}

A dynamic graph can be defined formally as an agglomerated graph $G=(V,E)$ and an ordered sequence of subgraphs $S=\{G_1=(V_1,E_1), G_2=(V_2,E_2), ..., G_k=(V_k,E_k)\}$ where each $G_t$ is the subgraph of $G$ at time $t$ where $t$ can be a specific time or it can be a time period. $V,V_1,V_2,...,V_k$ are finite and non-disjoint sets of nodes, $E,E_1,E_2,...,E_k$ are finite and non-disjoint sets of edges such that $V=V_1 \cup V_2 \cup ... \cup V_k$ and $E=E_1 \cup E_2 \cup ... \cup E_k$.

Other representations in literature only associate temporal dimension with edges and not with nodes (see for example \cite{Casteigts_2011}). With the described representation, it is  possible to represent when an object joins a system, and when it leaves it. This information might be very useful in some scenarios.

\subsection{Classifications of Dynamic Networks}

Dynamic networks can be classified in a number of ways. We do not try to provide an exhaustive list of classification methods, but merely describe some of the most commonly used discriminators from the literature. These classification methods are briefly described below: 

\subsubsection{Online versus Offline dynamic networks}

Online dynamic networks are the networks where we have streaming data and we have to analyse and visualize these networks with the arrival of data streams such as climate changes. Offline networks are the networks where the complete time-stamped data is available for analysis and visualization. For example, financial transactions of a banking system or an ATM machine.

\subsubsection{Continuous versus Discrete dynamic networks}

Discrete dynamic networks are networks where we have discrete time windows within which nodes appear and/or interactions take place for instance, the network of players in a football game. Continuous dynamic networks do not have any such discretization and continuous time scales are used to represent nodes and edges. For example, data packets moving over a local area computer network or the internet.

\subsubsection{Time-Unweighed versus Time-weighted dynamic networks}

Time-Unweighed dynamic networks, also known as Contact Sequence as defined by \cite{holme12}, are networks where interaction time among nodes is negligible and we only consider that an interaction took place such as an email. On the other hand Time-Weighted dynamic networks are those networks where interaction time is important. It represents some sort of weight that plays a vital role in the analysis and visualization of these networks. For example the duration of a phone call or the time it took to transfer data from one location to the other.


\subsection{Dynamic Graph Analytics}

Analysis of dynamic graphs requires methods and metrics that can cater to the temporal dimension of these networks. The most common method to deal with the temporal aspect is to simply agglomerate the entire network over time into a single static network, but this loses almost all temporal nuances.  Another common method used as preprocessing is the discretization of dynamic graphs: Given a temporal graph $G$ in a time period $t_{1,2}$ to $t_{n-1,n}$, discretization breaks $G$ into $n$ subgraphs $G_{1}, G_{2}, \cdots, G_{n}$ such that $G_{1}$ represents the state of the graph in time period $t_{1,2}$, $G_{2}$ represents the structure in time period $t_{2,3}$ and so on. We use the term ``time-step" to refer to individual subgraphs according to time periods sampled from $G$ and ``time window'' to refer to the time period for each subgraph. Such discretization of the network into a series of static networks can enable the extension of many static graph analysis techniques to be applicable to dynamic networks.

There are many structural analysis where the time ordering is an added constraint.  For example transitivity, i.e. the property where three or more nodes are all connected to each other. This property may not exist in a network if the edges connecting the nodes do not appear in the same time window.  However, this nuance would be lost in an agglomerate approach.

It is also possible to extend the time-step definition to a sliding time-window approach, where the discretization process for each $G_{a}$ with time period $t_{p,q}$ and $G_{a+1}$ with time period $t_{r,s}$ have some overlapping structure where the condition $r<q$ holds. This can serve to smooth out sharp changes in the network, and help in the isolation of outlying abnormalities.

\subsubsection{Dynamic Graph Models}

Some graph generating models have been proposed in the literature to mimic the dynamic processes taking place in real world networks. Notable contributions include the work by Robins et al. \cite{robins07} and a more recent model by Kolar et al. \cite{kolar10}. Often topological characteristics of real world networks are extracted and used as parameters for network models. The study of these models help us understand and predict how networks develop and change undergoing dynamic processes. Moreover benchmark datasets can be generated for testing algorithms and data analysis techniques for extensive evaluation.

\subsubsection{Dynamic Network Metrics}

Typically SNA metrics can be divided into element level (node or edge), group level (group of nodes) and network level metrics\cite{brandes05}.   
Traditional SNA metrics usually cannot be mapped directly to Dynamic networks due to the addition of the temporal dimension. Different researchers have proposed modifications to existing static network metrics or proposed new metrics altogether to analyze dynamic networks. Even fundamental network metrics such as `degree' need modification, as the number of connections for a node across different time periods may vary substantially. Thus new semantics must be associated with existing network metrics. 

One option that is often applied to create a static agglomerate network out of the entire dynamic network, weighted according to node/edge occurrence, which enables the applicability of SNA metrics.  However, this ignores the dynamic nature of the network.  Alternate weightings can be applied, such as weighting according to how `stable' each node/link is (i.e. how often they appear/disappear, not just how often they are there), but the result is still inherently static.

Another simple option is to simply recalculate metrics on static snapshots of the network (i.e. each time step).  This can then be analyzed using time-series analysis techniques.  However, it can still be na\"ive to analyze complex temporal patterns.

One interesting set of metrics for dynamic networks is concerned with paths and interconnectedness over time. Paths for static networks are sequences of nodes and edges such that the nodes are adjacent. In temporal graphs, instead of simple paths, the concept of time preserving paths is introduced such that a path can only exist from one node to the other only and only if the edges appear in non-decreasing order. Mathematically, for a path $E=\{e_{1} , e_{2} , \cdots ,  e_{n}\}$ the condition holds ($e_{t_{1}} \leq e_{t_{2}}, \cdots, e_{t_{n}}$) where $e_{t_{i}}$ denotes the time period associated with edge $e_{i}$. A simple implication of time preserving paths is that paths are no longer symmetric as opposed to static networks i.e.\ if a path from node $p$ to $q$ exists, a path from $q$ to $p$ might not necessarily exist. More details about temporal metrics can be found in \cite{holme12}.

\subsubsection{Dynamic Community Detection}



Static graph clustering has been an active area of research with well established methods and algorithms. Schaeffer \cite{Schaeffer2007} provides a good overview of some of the graph clustering methods. The purpose of such methods is to discover groups of nodes based on some similarity, either structural or based on some metric. Another successful set of methods helps to discover clusters of densely connected communities. They are generally based on an algorithm that optimizes a function such as the so-called modularity function from \cite{Newman2004} which represents the sum of the number of edges linking nodes of the same clusters minus the expected such sum if edges were distributed at random.

%
%


With the emergence of dynamic networks, clustering methods for static graphs fail to satisfy the new and challenging issues revolving around dynamic networks. Thus, a new area of research focuses on evolving communities within dynamic graphs: the underlying idea is to extract time-varying clusters (\textit{i.e.} clusters of densely connected communities that evolve over time), by extending or adapting the algorithms developed for static graphs.

Mathematically, we define dynamic graph clustering as follows: Let $G=(V,E)$ be a dynamic graph and $S=\{G_1=(V_1,E_1), G_2=(V_2,E_2), ..., G_k=(V_k,E_k)\}$ the corresponding ordered sequence of subgraphs. Following \cite{Sallaberry_2012}, a time-varying clustering of a dynamic graph $G$ is defined as a set of time-varying clusters $VC=\{VC_1,VC_2,...,VC_{l}\}$. Each of these time-varying clusters is an ordered sequence $VC_i=\{vc^1_i,vc^2_i,...,vc^k_i\}$ where $k$ is the number of time steps and each $vc^t_i$ is a subset of the vertices $V_t$ at time $t$. That is, each time-varying cluster $VC_i$ is a cluster whose membership can evolve over time, where $vc^t_i$ represents the set of nodes in the cluster $i$ at time $t$.

A first approach for creating a time-varying clustering is to apply directly a static graph clustering algorithm to an agglomerate of a dynamic graph $G=(V,E)$ \cite{Hu_2012}. It gives a partition $C=\{c_1,c_2,...,c_{l}\}$ of the nodes of $V$, which is used to create the dynamic clustering with $VC_i \leftarrow c_i$ for each time step and each $vc^t_i \leftarrow c_i \cap V_t$. As densely connected communities are extracted on the union of 
the time-steps, it does not guarantee that each cluster of each time-step is densely connected. 

An alternate approach to overcome this issue is to use a static graph clustering algorithm for each time-step subgraph $G_i$ and then associate the clusters accross time to derive time-varying clusters \cite{Sallaberry_2012}. In this approach, clusters are iteratively computed by comparing each time-step cluster in the current time-step pairwise with the time-step clusters of the previous time-step according to a similarity index and then greedily associating the time-step clusters that most closely match into the same time-varying cluster.  Once all clusters are assigned or the similarity falls below a user-defined threshold, this process terminates and moves to the next timestep. Any remaining new clusters that do not have a good enough match are considered new clusters, so they start new time-varying clusters. And any remaining time-varying clusters present in the previous time-step that were not assigned a cluster in the current time-step were discarded, as there was no match.

Cazebet \textit{et al.}\cite{cazabet10} also propose an interesting approach to detect dynamic communities. The algorithm is designed to 
detect strongly overlapping communities. The authors introduce two notions of intrinsic communities and longitudinal
detection which drives the iLCD (intrinsic Longitudinal Community Detection) algorithm.



\subsection{Dynamic Graph Visualizations}

Visualization methods for dynamic graphs present a challenging task for researchers. Traditional methods for the visualization of static graphs cannot be applied to dynamic networks. The problem is fundamental to the static representations used as they require at least two dimensions to represent proximity, as a result, time cannot be represented on a two dimensional plan \cite{moody05}. A simple approach to visualize a dynamic graph would have been as a static image but images fail to represent change occurring in networks over time \cite{moody05}.

Despite the success of visualization in static networks \cite{freeman00}, most of the approaches used in dynamic network analysis avoid using visualizations \cite{trier08}. These approaches mostly rely on measures and metrics to establish hypothesis and base their analysis on these statistics. A major reason for this limited use of visualizations is to ensure the stability of the layouts used. A stable layout helps preserve the user's mental map as there is less movement between time-steps, but sacrifices quality in terms of readability for later time-steps as their layout depends on previous time-steps. Many experiments have been proposed to examine the effect of preserving the mental map in dynamic graphs visualization \cite{Purchase_2008}. The results of \cite{Purchase_2008} were quite surprising because the most effective visualizations were the extreme ones, \textit{i.e.} the ones with very low or high mental map preservation: visualizations with medium preservation performed less well.


Dynamic network visualization helps to visualize dynamic network processes which in turn helps to build models that can predict future evolutionary processes. Static representation fails to accurately build a dynamic model because using graphs, we develop a static model of a phenomenon which is inherently dynamic, and then we try to make inferences and predictions of how it will evolve over time. 

Visualization methods for dynamic graphs can be categorized into animated visualizations, static visualizations and hybrid visualizations which are described below.

\subsubsection{Animated visualizations} 

A common method for visualizing dynamic graphs is to animate the transitions between time-steps \cite{Frishman_2008}. For example, the approaches proposed by \cite{moody05,Frishman_2008} render individual network graphs segmented from entire dynamic graph. These individual graphs are then visualized as animated sequence with nodes appearing, disappearing and moving to produce readable layout for each time-step. This animation facilitates the analysis and understanding of dynamic processes taking place in a network and thus has gained a lot of popularity. One drawback of this approach is its scalability as number of nodes and processes increases, it becomes difficult to visually comprehend the changes taking place in the network.

Approaches to visualize dynamic clustered graphs also exist in the literature such as Hu et al.\ \cite{Hu_2012} who proposed a method based on a geographical metaphor to visualize clustered dynamic graphs. 

All the aesthetic requirements for an animated visualization remains the same as that of static visualization such as minimizing edge crossings and node overlap. One aspect which needs to be handled for animations is the movement of nodes between time-steps which needs to be meaningful and coherent with the network processes taking place. 

\subsubsection{Static visualizations} 

Static visualizations have also been successfully used to visualize dynamic networks. Different techniques have been proposed in the literature such as ``Small multiples'' \cite{Tufte_1990} where snapshots of different time-steps are placed next to each other. This technique eases the comparison of distant time-steps but the area devoted for each time-step is small and this reduces the readability of each subgraph. Another technique named `Flipbook' \cite{moody05} only displays edges within a time window whereas nodes maintain a fixed position. As the time window is moved, edges appear and disappear showing interactions taking place at different time intervals.

\begin{figure}[!h]
  \centering
  \subfloat[]{\label{fig:sm}\includegraphics[height=0.39\textwidth]{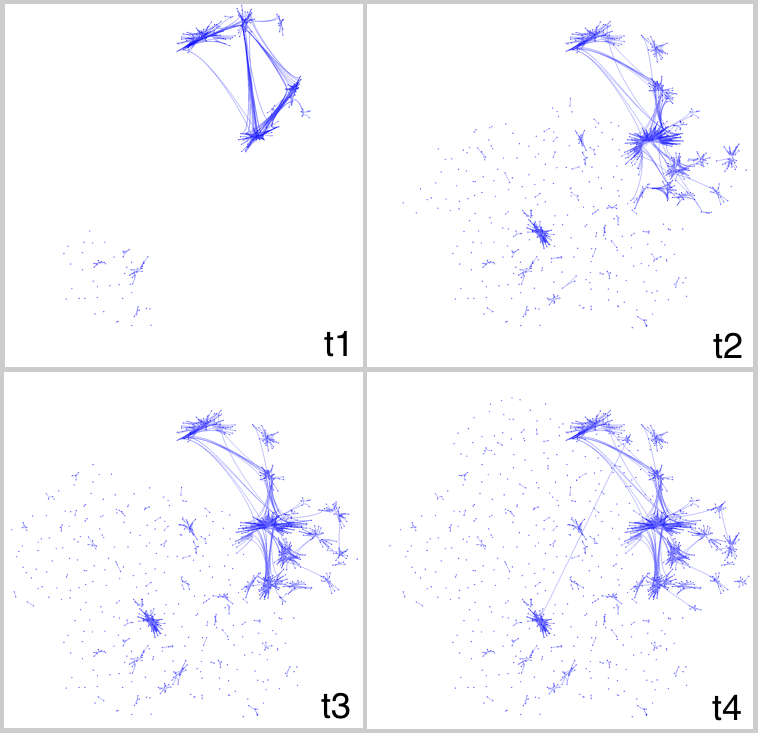}}                
  \hspace{0.3cm}
  \subfloat[]{\label{fig:burch}\includegraphics[height=0.39\textwidth]{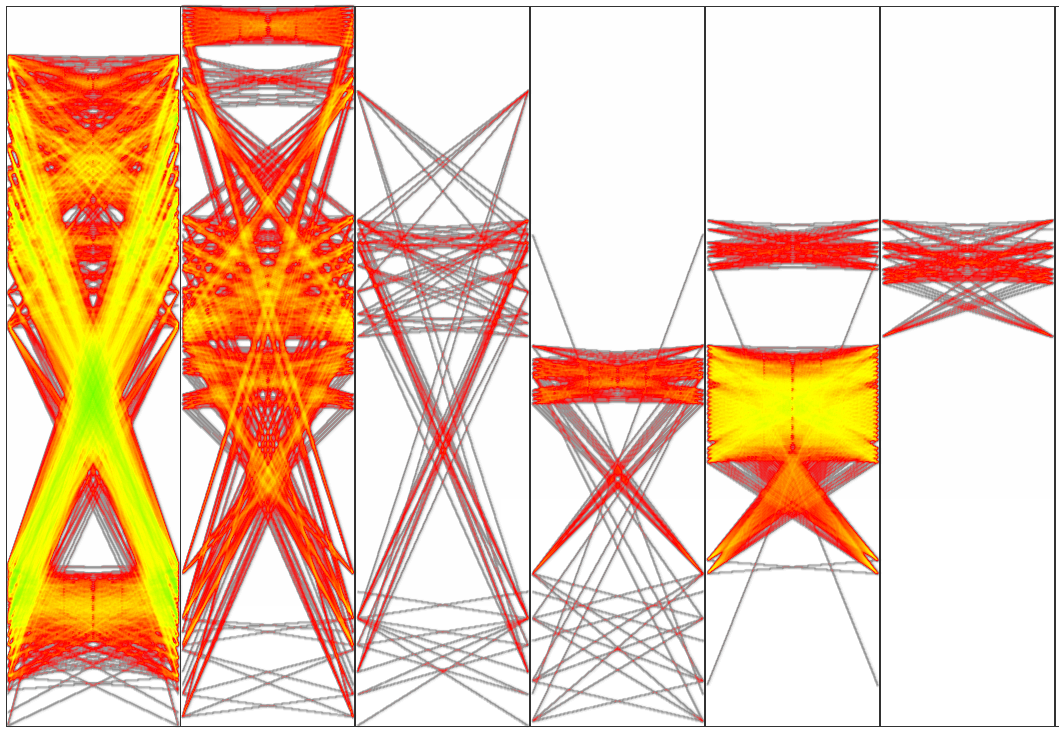}}
  \caption{(a) Small multiples: snapshots of different time-steps are placed next to each other. (b) Parallel edge splatting for scalable dynamic graph visualization: vertices are ordered and positioned on several vertical parallel lines, and directed edges connect these vertices from left to right (image from \cite{Burch_2011} and reproduced with permission).}
  \label{fig:static}
\end{figure}

Another interesting visualization technique dealing with dynamic large directed graphs has been proposed by Burch \textit{et al.} \cite{Burch_2011} where vertices are ordered and positioned on several vertical parallel lines, and directed edges connect these vertices from left to right. Each time-step's graph is thus displayed between two consecutive vertical axes (see figure \ref{fig:burch}).

\subsubsection{Hybrid Visualizations}

\begin{figure}[!h]
  \centering
  \subfloat[]{\label{fig:saltl}\includegraphics[width=0.37\textwidth]{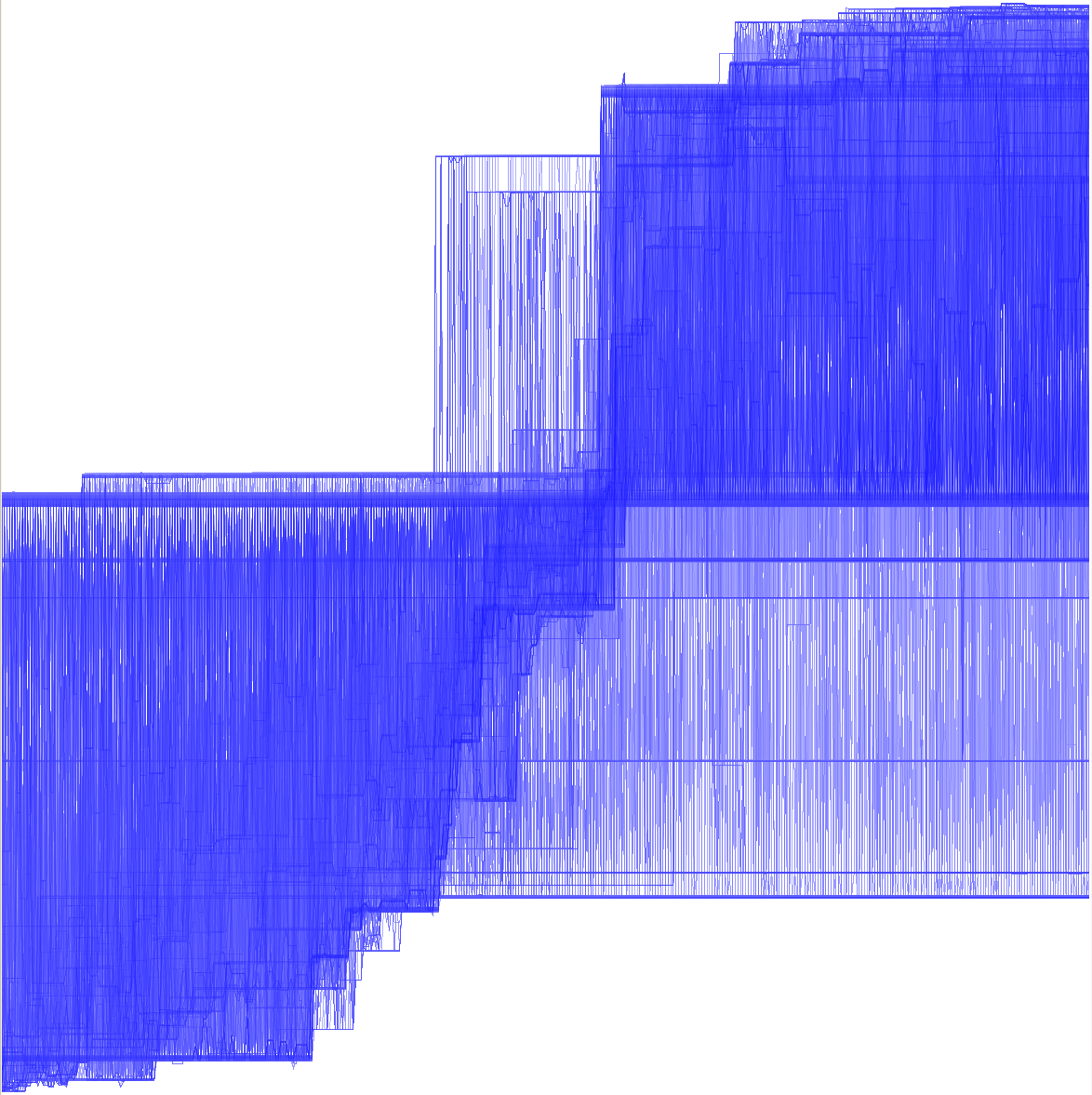}}                
  \hspace{1cm}
  \subfloat[]{\label{fig:salgr}\includegraphics[width=0.45\textwidth]{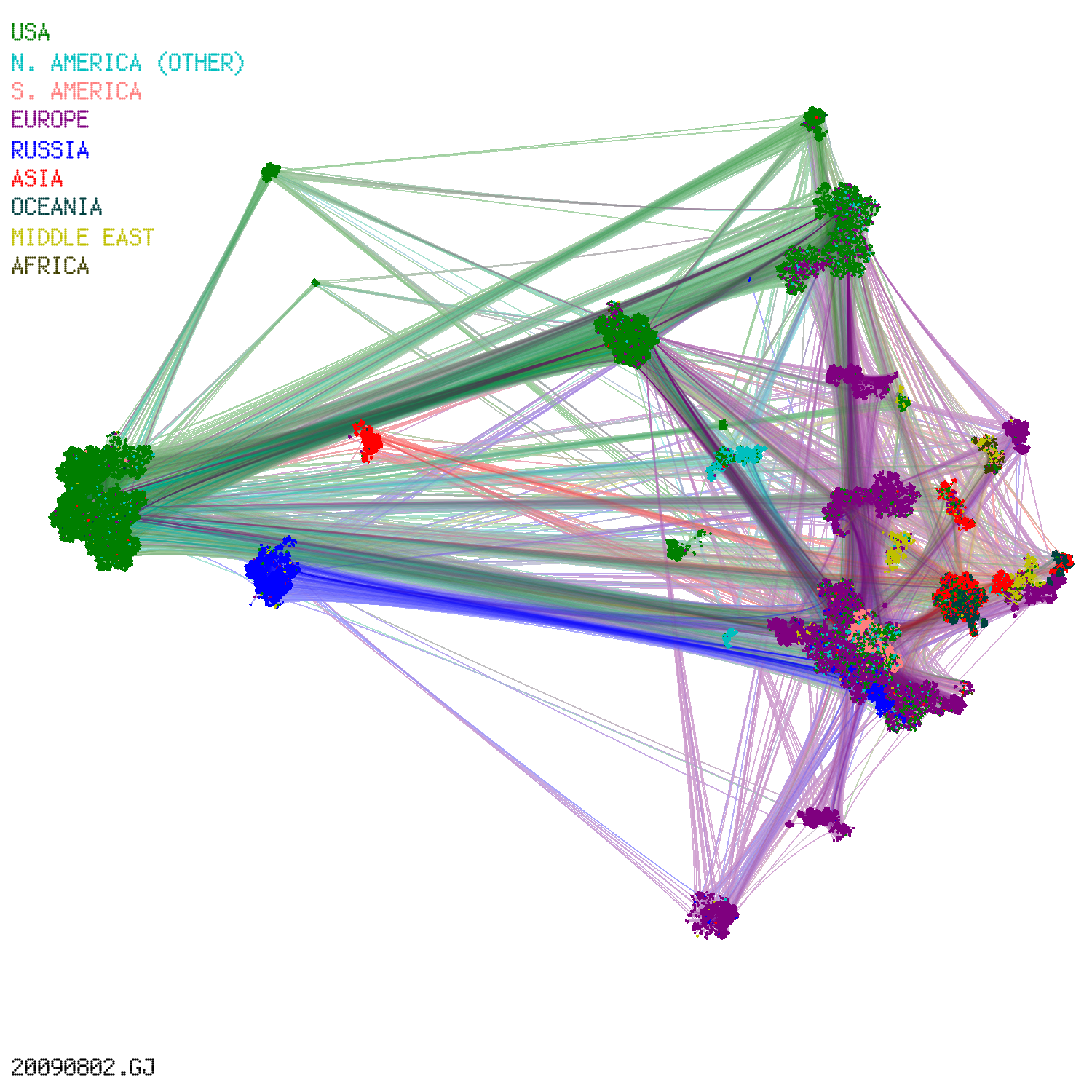}}
  \caption{(a) Overview: each node is represented as a line where the x-position is time and the y-positions corresponds the cluster the nodes belong to at each given time and its position within the cluster. (b) Time-step view: node-link diagram that shows the graph at any selected time-step.}
  \label{fig:hybrid}
\end{figure}

Some approaches combine both static representations with dynamic layouts to provide both a summary overview as well as detailed views.  One such method\cite{Sallaberry_2012} computes time-varying clusters and orders both the clusters and individual nodes in 1 dimension.  This is used to both define a 2-dimensional overview of the network over time as well as defining temporally stable layouts for any given time (see figure \ref{fig:hybrid}).

\section{Key Applications}

\textbf{Biological Networks:} One of the most successful applications of DNA is in the area of biological networks. There are a number of visual analysis tools such as Cytoscape, VisANT, Pathway Studio, Patika and ProVis-Tulip \cite{suderman07,pavlopoulos08}. These tools currently focus on the dynamic behaviour of biological systems like modeling cellular processes, subcellular localization information and time-dependent behaviour, interaction of protein-protein/protein-nucleotide networks \cite{suderman07,akhmanova08}. This recent interest is due to the huge amount of dynamic data available where visualization tools have limited capabilities to present readable images from these networks. Thus the dynamism of these networks allow domain experts to focus on small time periods with minimal activity making it easier for analysis and drawing conclusions.

For example, Akhmanova and Steinmetz  \cite{akhmanova08} study microtubule plus-end tracking proteins which  form dynamic networks through the interaction of a limited set of proteins modules. Taylor et al \cite{taylor09} study the dynamic structure of human protein interaction network to predict the presence of breast cancer using structural changes.

\textbf{Terrorist Networks:} Another highly sensitive area where DNA is widely utilized, is the area of security and terrorist networks. Criminals and terrorists work as a cohesive group to achieve desired outcomes and no single person is responsible to do it all. The roles and duties of these individuals keep changing as well as new individuals are introduced in the network regularly. Furthermore lot of information is missing or inaccurate making it a challenging domain for social network researchers. A famous success story of the application of this approach is the capture of former president of Iraq, Saddam Hussein \cite{borgatti09}.

One notable work is the software system named DynNetSim by Adler \cite{adler07} to model and analyze dynamic networks. The software provides a holistic view of networks responding to influences of environmental forces and disruptive events. DynNetSim also helps to study impacts of strategies to change networks and enhance different counter-terrorism activities. Gilbert et al \cite{gilbert11} studied the dynamic behaviour of terrorist groups given the data of cellular phones. They provide a complete framework starting from time-stamped network data to presenting visual drawings for analysis and knowledge extraction.

\textbf{Computer Networks:}  Computer security often requires monitoring of complex and highly dynamic networks, often in real-time. There are many layers to computer networks, including physical connectivity, routing tables, network flows/traces, or even application level semantics such as online social networks or hyperlinks on the World Wide Web.  There are numerous approaches that have been applied to these networks.  Due to the highly dynamic nature of some of these networks (e.g. network flows), many approaches ignore the topology.  However, some levels of this network are temporally stable enough to apply DNA (e.g. routing tables or hyperlinks).  One example particularly worth noting here is \cite{Sallaberry_2012}, which investigates routing tables at the scale of the entire Internet.

%
%

\section{Conclusion and Future Directions}

A major reason for the popularity of the field of dynamic networks is its applicability in a number of diverse fields. The field of dynamic networks is in its infancy and there are so many avenues that need to be explored. From developing network generation models to developing temporal metrics and measures, from structural analysis to visual analysis, there is room for further exploration in almost every dimension where dynamic networks are studied. Recently, with the availability of dynamic data from various fields, the empirical study and experimentation with real data sets has also helped maturate the field. Furthermore, researchers have started to develop foundations and theories based on these datasets which in turn has resulted lots of activity among research communities.

While there is a growing corpus of works on dynamic graph analysis, there are still many directions for further investigation.  Dynamic graph metrics are largely unexplored, as most existing dynamic graph metrics are merely applications of static graph metrics.  Likewise, dynamic graph clustering is still in its infancy, as most effective dynamic graph clusterings are currently based on direct extensions to static graph clusterings.  While dynamic graph visualization methods have been establishing a growing foothold, there is still much room for novel approaches.  In particular, scalability will only become a more important issue, as there are very few works that can can handle dynamic networks at the scale of many of today's large real-world networks.  Addressing topics such as these will serve to greatly further both the field of network analysis and any other of the vastly numerous field where such dynamic networks occur.






%

\section{Cross-References}

This book contains many articles focusing on related topics such as network analysis, community detection, etc. We have listed below some of the most relevant titles: 
\begin{itemize}
  \item 00223 - Community Evolution,
  \item 00215 - Community Discovery and Analysis in Large-Scale Online/Offline Social Networks,
  \item 00380 - Community Identification in Dynamic and Complex Networks,
  \item 00031 - Large Networks, Analysis of
\end{itemize}

This book also includes several papers dealing with network visualization:
\begin{itemize}
  \item 00303 - Social Network Construction, Visualization and Analysis Tools: Introduction,
  \item 00268 - Visual Methods for Social Network Analysis,
  \item 00044 - Visualization of Large Networks.
\end{itemize}

Indeed, some visualization and network analysis softwares are described in these articles: 
\begin{itemize}
  \item 00299 - Gephi,
  \item 00304 - JUNG,
  \item 00315 - Tulip,
  \item 00316 - UCINET,
  \item 00310 - Pajek,
  \item 00309 - ORA.
\end{itemize}

\bibliographystyle{spbasic}
\bibliography{biblio}

\section{Recommended Reading}

The current literature lacks a comprehensive text covering all aspects of dynamics networks. A highly informative and recent work titled `Temporal Networks' by Holme and Saram\"{a}ki \cite{holme12} reviews most of the analytical part related to dynamic networks from the current literature. The authors primarily focus on dynamic network metrics, methods of representing dynamic data as static networks, models for generating temporal networks and the spreading dynamics in these networks.

One of the landmark papers for dynamic network visualization is titled `Dynamic Network Visualization' by Moody et al.\cite{moody05} where they introduce two important concepts of network visualizaion, network movies and flipbook. Bender-deMoll and McFarland's \cite{demoll06} article `The Art and Science of Dynamic Network Visualization' is also very interesting to read as it reviews existing layout algorithms for static networks and how they can be used for visualization of dynamic networks. M. Trier \cite{trier08} also studies the problem of dynamic network visualization to analyze online social communities. The author uses animated graphs and measure changes to describe cluster formation processes, relate node level analysis to network level analysis and measure how external events change network structures.

\end{document}